# Photothermal heterodyne imaging of individual non-fluorescent nano-objects

Stéphane Berciaud, Laurent Cognet, Gerhard A. Blab, and Brahim Lounis*

*Centre de Physique Moléculaire Optique et Hertzienne, CNRS (UMR 5798) et Université Bordeaux I, 351, cours de la Libération, 33405 Talence Cedex, France*

We introduce a new, highly sensitive, and simple heterodyne optical method for imaging individual non-fluorescent nano-objects. A two orders of magnitude improvement of the signal is achieved compared to previous methods. This allows for the unprecedented detection of individual small absorptive objects such as metallic clusters (of 67 atoms) or non-luminescent semiconductor nanocrystals. The measured signals are in agreement with a calculation based on the scattering field theory from a photothermal-induced modulated index of refraction profile around the nanoparticle.

In the fast evolving field of nanoscience, where size is crucial for the properties of the objects, simple and sensitive methods for the detection and characterization of single nano-objects are needed. The most commonly used optical techniques are based on luminescence. Single fluorescent molecules have been studied on their own and are now routinely applied in various research domains ranging from quantum optics[1-3] to life science[4]. Yet, fluorescent molecules allow only for short observation times due to inherent photo-bleaching. The development of brighter and more stable luminescent objects, such as semiconductor nano-crystals[5, 6] has remedied some of this shortcoming but this improvement has come at the price of a strong blinking behavior.

An interesting alternative to fluorescence methods relies solely on the absorptive properties of the object. At liquid helium temperatures single molecules were initially detected by an absorption technique owing to the high quality factor of the zero-phonon line which gives a considerable absorption cross-section at resonance[7] (few $10^{-11}$ cm$^2$). Single ions or atoms isolated in rf-traps[8] or high Q cavities[9, 10] have been detected by absorption of a probe beam. In general, particles with large absorption cross sections and short time intervals between successive absorption events are likely candidates for detection with absorption methods[8].

Metal nanoparticles fulfill both of these requirements: excited near their plasmon resonance a nanometer sized gold nanoparticle has a relatively large absorption cross section (~8 $10^{-14}$ cm$^2$ for a 5 nm diameter particle) and a fast electron-phonon relaxation time (in the picosecond range[11]). Since luminescence from these particles is extremely weak, almost all the absorbed energy is converted into heat. The temperature rise induced by the heating leads to a variation of the local index of refraction. Previously, a polarization interference contrast technique has been developed[12] to detect this photothermal effect. In that case, the signal is caused by the phase-shift induced between the two spatially-separated beams of an interferometer, where only one of the beams propagates through the heated region, and images of 5nm diameter gold nanoparticles have been recorded with a signal-to-noise ratio SNR ~ 10. Also, the sensitivity of this technique, although high, is ultimately limited by the quality of the overlap of the two arms of the interferometer as well as by their relative phase fluctuations.

In this letter we introduce a new, more sensitive, and much simpler method for detecting non fluorescent nano-objects. It uses a single probe beam which produces a frequency-shifted scattered field as it interacts with time-modulated variations of the refraction index around an absorbing nano-object. The scattered field is detected by its beatnote with the probe field which plays the role of a local oscillator as in heterodyne technique. Because this new method is not subject to the limitations mentioned above, a two orders of magnitude improvement of the sensitivity is achieved compared to the previous photothermal method. This allows for the unprecedented detection of small absorptive objects such as individual metallic clusters composed of 67 atoms.

When a small gold nanoparticle (or any absorbing nano-object) embedded in a homogeneous medium is illuminated with an intensity modulated laser beam, it behaves like a heat point source with a heating power $P_{heat} \cdot [1 + \cos(\Omega t)]$, $\Omega$ being the modulation frequency and $P_{heat}$ the average absorbed laser power. It generates a time-modulated index of refraction in the vicinity of the particle with a spatio-temporal profile given by[13] $\Delta n(r,t) = \frac{\partial n}{\partial T} \frac{P_{heat}}{4\pi\kappa r}\left[1 + \cos\left(\Omega t - \frac{r}{R_{th}}\right)e^{-\frac{r}{R_{th}}}\right]$, where r is the distance from the particle, $n$ the index of refraction of the medium, $\frac{\partial n}{\partial T}$ its variations with temperature $(\sim 10^{-4} K^{-1})$, $R_{th} = \sqrt{2\kappa/\Omega C}$ the characteristic length for heat diffusion, $\kappa$ the thermal conductivity of the medium and C its heat capacity per unit volume. A probe beam interacting with this profile gives rise to a scattered field containing sidebands with frequency shifts $\Omega$. As any heterodyne technique, interference between a reference field $E_{ref}$ (either the reflection of the incident probe at the interface between a cover slip and the sample or its transmission, see figure 1A) and the scattered field produces a beatnote at the modulation frequency $\Omega$ which can be easily extracted with a lock-in amplifier.

In practice, we overlay a (red) probe beam (720nm, single frequency Ti:Sa laser) and a (green) heating beam (532 nm, frequency doubled Nd:YAG laser) whose intensity is modulated at $\Omega$ (100 kHz to 15 MHz) by an acousto-optic modulator (see figure 1A). Using a high aperture objective (100x, Zeiss, NA=1.4), both beams are focused onto a same spot on the sample. A combination of a polarizing cube and a quarter wave plate is used to extract the interfering reflected field (so-called reference field) and backward scattered field. Optionally, a second microscope objective can be employed to efficiently collect the interfering probe-transmitted and forward-scattered fields. The power of the heating beam ranged from less than 1μW to 3.5 mW (depending on the nanoparticle size to be imaged) at the objective. Reflected or transmitted red beams are collected on fast photodiodes and fed into a lock-in amplifier to detect the beat signal at $\Omega$. Throughout the experiment, we used an integration time of 10 ms. A microscopy image was formed by moving the sample over the fixed laser spots by means of a 2D piezo-scanner.

The samples were prepared by spin-coating a solution of gold nanoparticles (diameter of 1.4nm, 2nm, 5nm, 10nm, 20nm 33nm or 75nm, diluted into a polyvinyl-alcohol matrix, 2% mass PVOH) onto clean microscope cover slips. The dilution and spinning speed were chosen such that the final density of spheres in the sample was less than 1 $\mu m^{-2}$. Application of a silicon oil on the sample insures homogeneity of the heat diffusion. The size distribution of the nanospheres was checked by transmission electron microscopy (data not shown) and was in agreement with the manufacturer's specification.

Figure 1B shows a three-dimensional representation of a photothermal heterodyne image of small gold aggregates of 67 atoms (1.4 nm nanogold). The image shows no background from the substrate, which means that the signal arises from the only absorbing objects in the sample, namely the gold aggregates. They are detected with a relatively small heating power (~3.5mW) and a remarkably large signal-to-noise ratio (SNR >10). We further confirmed that the peaks stem from single particles by generating the histogram of the signal height for 272 imaged peaks (Figure 1C). We found a monomodal distribution with a width in agreement with the spread in particle size.

In order to estimate the measured signal, we have used the theory of "scattering from a fluctuating dielectric medium"[14] to calculate the field scattered by the modulated index profile $\Delta n(r,t)$. The beating at $\Omega$ between the reference and scattered fields leads to a beating power $S$ at the detector with two terms in quadrature[15]:

$$S = \alpha\, n\, \frac{\partial n}{\partial T} \sqrt{I_{inc}} \sqrt{P_{ref}} \frac{P_{heat}}{C\lambda^2} \frac{1}{\Omega} \left[ f_\kappa(\Omega)\cos(\Omega t) + g_\kappa(\Omega)\sin(\Omega t) \right] \quad (1)$$

with $\alpha$ a geometry factor close to unity, $I_{inc}$ the incident red intensity at the particle location and $P_{ref}$ the reference (back-reflected) beam power. $f_\kappa(\Omega)$ and $g_\kappa(\Omega)$ are two dimensionless functions which depend on the modulation frequency and the thermal diffusivity of the medium. The variations of $f_\kappa(\Omega)/\Omega$ and $g_\kappa(\Omega)/\Omega$ are presented on figure 2A for $\kappa/C=2.10^{-8} \text{m}^2/\text{s}$. At low frequencies, the characteristic length of the heat diffusion $R_{th}$ is larger than the probe spot size ($\sim\lambda/2$) and the term $f_\kappa(\Omega)/\Omega$, in phase with the applied modulation, is preponderant. However, at sufficiently high frequencies such that $R_{th} \ll \lambda$, the quadrature term $g_\kappa(\Omega)/\Omega$ dominates and decreases as $1/\Omega$.

The magnitude of demodulated signal delivered by the lock-in amplifier is proportional to:

$$S_{dem} \propto \sqrt{\left\langle S(t)^2 \right\rangle_t} \propto \frac{1}{\Omega} \sqrt{f_\kappa(\Omega)^2 + g_\kappa(\Omega)^2} \quad (2)$$

The frequency dependence of this signal measured on single particles is presented on figure 2. A good quantitative agreement with the theoretical form of $S_{dem}$ is obtained.

To further ensure the validity of our calculations, we used Eq. 1 to estimate the beating power at the detector. A single 2nm gold nanoparticle has an absorption cross section of $\sim 5.10^{-15}$ cm$^2$ at 532nm and will absorb $P_{heat} = 10 nW$ when illuminated by a laser intensity of 2MW/cm$^2$. For a probe incident power $P_{inc}$=70mW, a frequency $\Omega/2\pi$ =800kHz and a reference power $P_{ref}$=100µW, Eq. 2 gives the beating power $S_{dem}$~5nW. After calibration of the detection chain, we measured a beating power of ~2nW in qualitative agreement with the theoretical prediction.

Figure 3B shows a linear dependence of the signal with heating power. A further increase on the power is not accompanied by saturation but leads to fluctuations in the signal amplitude and eventually irreversible damage on the particle[16, 17].

As a first application of this method, we studied the size dependence of the absorption cross section of gold nanoparticles (at 532nm, close to the maximum of the plasmon resonance) with diameters ranging from 1.4nm to 75 nm. To do so, we prepared different samples containing nanoparticles of two different (successive) sizes (1.4 & 5nm, 2 & 5 nm, 5 & 10 nm up to 33 & 75 nm). For each sample, a histogram of the signal amplitudes was generated as exemplified on figure 3. All the histograms displayed bimodal distributions and the mean of each population was measured. This allows to report the size dependence of the absorption cross section normalized to that of 10 nm particles (figure 3B). As expected by the Mie scattering theory, we find a good qualitative agreement with a third-order law of the absorption cross-section vs the radius of the particles[17] (solid line on figure 3B). By removing ensemble averaging, this approach opens up the possibility to investigate the size dependent optical material functions of small metal clusters such as the dielectric permittivity[18].

As shown in figure 1, we are now able to detect metal nanoparticles as small as 1.4 nm in diameter with a good SNR (>10) which is shot-noise limited. To our knowledge, this is the first time that such small aggregates are being detected with purely optical methods. While the absorption cross-section of these clusters is only of the order of 10$^{-15}$ cm$^2$, comparable to that of a good fluorophore, or CdSe/ZnS nanocrystals[19, 20], their relaxation times are very short. In contrast, luminescent semiconductor nanocrystals or fluorescent molecules have radiative relaxation times in the nanosecond range, which renders them difficult to be detected by their absorption. However, at relatively high excitation intensities, semiconductor nanocrystals do no longer exhibit luminescence. Efficient non-radiative relaxation pathways open up with short relaxation times relying on Auger multi-exciton relaxation processes[21, 22] which make them detectable by our method. Figure 4a shows a fluorescent image of luminescent colloidal CdSe/ZnS quantum dots (peak emission at 640nm) excited by the heating beam at very low intensities (0.1 kW/cm$^2$). The blinking behavior characteristic of single quantum dot emission, is clearly visible in the image[23]. A photothermal heterodyne image of the same region was recorded afterwards at an excitation intensity of 5MW/cm$^2$ where quantum dots are no longer luminescent (figure4B). The two images correlate well, ensuring that the spots in the photothermal heterodyne image are indeed individual quantum dots (>90% of the fluorescent spots correlate with a photothermal spot). They do not show any blinking behavior. Interestingly, initially non-fluorescent quantum dots (absent from Figure 4A) are now detected by the photothermal technique.

For biological applications, the temperature rise at the surface of the nanoparticle is an important issue[24]. In the current configuration, a 5 nm gold nanoparticle can be detected with a SNR >100 at a heating power of 1mW. At this power, we estimate a local temperature increase of 4 K in aqueous solutions. As the temperature reduces as the inverse of distance, and most conceivable microscopy applications in biosciences do not require such a high SNR, the method presented in this letter will permit to image small gold particles by inducing a local heating of far less than 1 K above the average temperature in the sample.

The present work demonstrates the advantages of photo-thermal heterodyne detection for absorbing nano-objects. As any far-field optical technique, it has a wavelength limited resolution. An interesting challenge would now be to combined the unprecedented sensitivity of the method presented here with the sub-wavelength resolution of near-field optical techniques[25]. The study of the physical properties of very small metallic aggregates or non-luminescent semiconductor nanocrystals is now possible at the individual object level. This photothermal method doesn't suffer from the drawbacks of blinking and photobleaching and is immune to the effects of fluorescing and scattering backgrounds. It could be applied to many diffusion and co-localization problems in physical chemistry and material science and to track labeled bio-molecules in cells.

**Acknowledgement**

We wish to thank A. Brisson and O. Lambert for their assistance with electron microscopy experiments, P. Tamarat and O. Labeau for their help with the quantum dots, M. Orrit and D. Choquet for helpful discussions,. GAB acknowledges financial support by FWF (Schrödinger-Stipendium) and the Fondation pour la Recherche Médicale. This research was funded by CNRS (ACI Nanoscience and DRAB), Région Aquitaine and the French Ministry for Education and Research (MENRT).

25    A. Hartschuh, E.J. Sanchez, X.S. Xie, *et al*., Phys. Rev. Lett. **90**, 09503 (2003).

**Figure captions:**

**Figure 1:** (A) Schematic of the experiment. (B) 3D representation of a photothermal heterodyne image (5x5 μm$^2$) containing individual 67 atoms gold-clusters (1.4 nm diameter). (C) Signal histogram of 272 peaks detected in a sample prepared with 67 atoms gold-clusters. The monomodal shape of the distribution reveals that individual clusters are detected.

**Figure 2:** (A) Measured dependence of the signal (diamonds) on the modulation frequency $\Omega$ measured on a single 5 nm nanoparticle and comparison with theory using equation (2) (solid line). The variations of $f_\kappa(\Omega)/\Omega$ (dash) and $g_\kappa(\Omega)/\Omega$ (dash-dot) are also presented.
(B) Signal obtained from an individual 5 nm gold nanoparticle (squares) as a function of the incident power. The data are adjusted by a linear fit (solid line).

**Figure 3:** (A) Signal distribution obtained from a sample containing both 2 nm and 5 nm gold nanoparticles. (B): size dependence of the signal i.e. absorption cross section (circles) deduced from a series of histograms as presented in Figure 3 A, and comparison to the Mie theory (solid line)

**Figure 4:** Comparison of luminescence (A) and photothermal (B) images of the same area in a sample containing CdSe/ZnSe semiconductor nanocrystals. The insets show a zoom of one individual quantum dot marked by a square in the lower right of each image. Scale bar is 1μm.

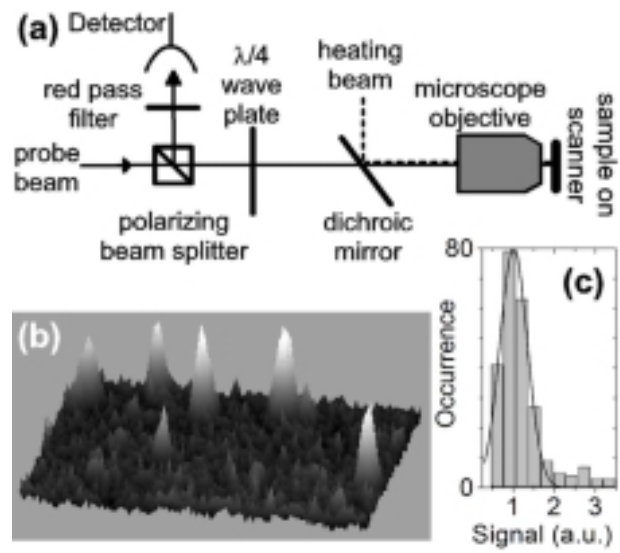

**Figure 1**

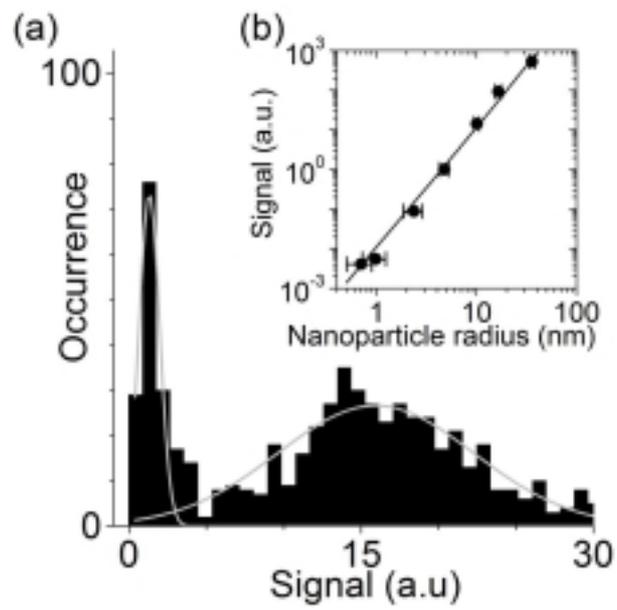

**Figure 2**

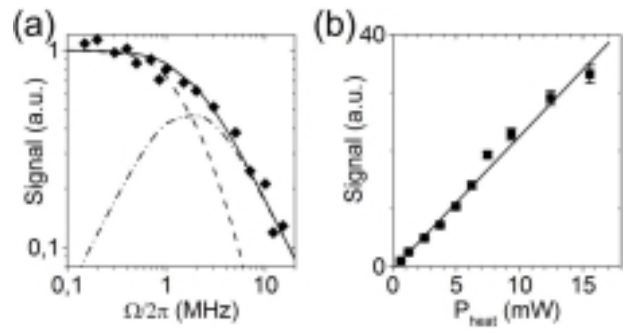

**Figure 3**

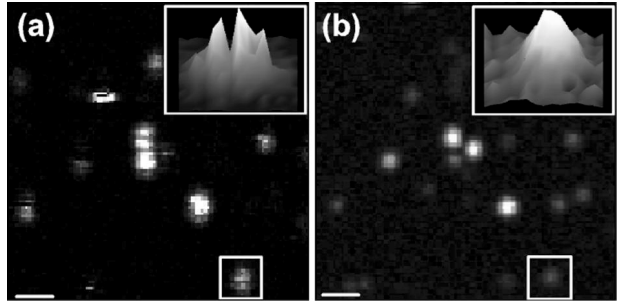

**Figure 4**